\RequirePackage{fix-cm}
\documentclass[smallextended]{svjour3}                     
\smartqed  

\usepackage[margin=1.5in]{geometry}
\usepackage{graphicx}
\usepackage{setspace}
\usepackage{lineno}
\linenumbers

\usepackage{natbib}
\usepackage{mhchem}
\usepackage{times,mathptmx}
\usepackage{balance} 

\usepackage{lastpage}
\usepackage[format=plain,justification=raggedright,singlelinecheck=false,font=small,labelfont=bf,labelsep=space]{caption} 
\usepackage{fancyhdr}
\usepackage{gensymb}
\usepackage{amssymb}
\usepackage{epsfig}
\usepackage{subfig}
\usepackage{bm}
\usepackage{wasysym}
\usepackage{pxfonts}
\usepackage{textcomp}
\usepackage{pxfonts}
\usepackage{color}
\usepackage{epstopdf}
\begin{document}

\title{Diffusion NMR Study of Complex Formation in Membrane-Associated Peptides}


\author{Suliman Barhoum, Valerie Booth and Anand Yethiraj         
}


\institute{ Suliman Barhoum \at
              Department of Physics and Physical Oceanography, Memorial University of Newfoundland, St. John's, NL, Canada \\
              \email{sulimanb@mun.ca},\\ 
           Valerie Booth \at
                 	 Department of Biochemistry,  Memorial University of Newfoundland, St. John's, NL, Canada \\
                 	 Tel.: 709-864-4523\\
                 	 \email{vbooth@mun.ca},\\    
           Anand Yethiraj \at
              Department of Physics and Physical Oceanography, Memorial University of Newfoundland, St. John's, NL, Canada \\
              Tel.: 709-864-2113\\
              \email{ayethiraj@mun.ca}     }
\date{Received: date / Accepted: date}


\maketitle


\begin{abstract}

Pulsed-field-gradient nuclear magnetic resonance (PFG-NMR) is used to obtain the true hydrodynamic size of complexes of peptides with sodium dodecyl sulfate SDS micelles. The peptide used in this study is a 19-residue antimicrobial peptide, GAD-2. Two smaller dipeptides, alanine-glycine (Ala-Gly) and tyrosine-leucine (Tyr-Leu), are used for comparison. We use PFG-NMR to simultaneously measure diffusion coefficients of both peptide and surfactant. These two inputs, as a function of SDS concentration, are then fit to a simple two species model that neglects hydrodynamic interactions between complexes. From this we obtain the fraction of free SDS, and the hydrodynamic size of complexes in a GAD-2--SDS system as a function of SDS concentration. These results are compared to those for smaller dipeptides and for peptide-free solutions. At low SDS concentrations ([SDS] $\leq$ 25 mM), the results self-consistently point to a GAD-2--SDS complex of fixed hydrodynamic size R =(5.5 $\pm$ 0.3) nm. At intermediate SDS concentrations (25 mM $<$ [SDS] $<$ 60 mM), the apparent size of a GAD-2--SDS complex shows almost a factor of two increase without a significant change in surfactant-to-peptide ratio within a complex, most likely implying an increase in the number of peptides in a complex. For peptide-free solutions, the self-diffusion coefficients of SDS with and without buffer are significantly different at low SDS concentrations but merge above [SDS]=60 mM. We find that in order to obtain unambiguous information about the hydrodynamic size of a peptide-surfactant complex from diffusion measurements, experiments must be carried out at or below [SDS] = 25 mM.

\end{abstract}

\section*{Introduction}
\label{intro}
Membrane-associated proteins and peptides are often studied in a micellar environment \citep{Tulumello2009,Charles2006}. Like membrane bilayers, micelles provide a hydrophobic-hydrophilic interface, but unlike them, they are small enough to enable solution NMR signals to be observed. Micelles are commonly employed in NMR structure determination of membrane proteins \citep{Goto2012,Tulumello2009}, but have also been used in studies where the protein-lipid interaction itself is the focus \citep{SaraCozzolino2008,SvenMorein1996,LanlanYu2006,AnaPaulaRomani2010}. NMR-based techniques have been utilized to study an important class of membrane-associated proteins that are called antimicrobial peptides (AMPs).

AMPs are often short peptides consisting of 12 to 50 residues and act by interacting with (and often disrupting) membranes. AMPs have been shown to play an important role in attacking and killing microbes such as bacteria, viruses, and fungi \citep{Zasllof2002,Nicolas2009,Hoskin2008,Chinchar2004}. Moreover, some AMPs exhibit activity against tumor cells in a mammal's body by disrupting the membrane of the diseased cells and targeting the cell interior without affecting the membrane of host cells \citep{Rege2007}. This selectivity, for microbial and/or tumor cells, is thought to arise due to the amphiphilic structure of the AMP that has an affinity to the lipid bilayer structure of the microbial cells as well as due to the interaction between the positive charge on the AMP with the anionic components of the tumor or pathogen cell membrane \citep{Epand1999}. Therefore, anionic sodium dodecyl sulfate SDS surfactant micelles are commonly employed in the structural studies of AMPs, as well as other membrane proteins \citep{Guangshun2008,Guangshun2006,TracyWhitehead2001,Orfi1998,Begotka2006,Deaton2001,TracyWhitehead2004,Gao1998,Buchk1998}.   

A knowledge of the hydrodynamic size of proteins plays an important role in understanding their conformations \citep{Dobson1997}. This is also the case for peptides in peptide-micelle complexes, where there could be many coexisting conformations. The hydrodynamic size of complexes can be obtained by measuring diffusion coefficients and using the Stokes-Einstein-Sutherland equation $\mathrm{R_{H}=K_{B}T/6 \pi \eta D_{o}}$. This approach, however, is only strictly valid when the self-diffusion coefficient $\mathrm{D_{o}}$ is obtained by measuring the diffusion coefficient as a function of the surfactant concentration and then extrapolating to infinite dilution. Such a procedure is often not practical when the amount of peptide or protein is limited in quantity. As a result of this, ``apparent'' hydrodynamic radii are routinely reported, without such extrapolation, in systems with rather large surfactant concentrations \citep{Waks1989, Brown1996, Booth2011}.
 
An important phenomenon to consider with respect to large macromolecular concentrations is crowding. Macromolecular crowding usually refers to the non-specific excluded volume (steric) effect of macromolecules with respect to one another in an environment where the macromolecular volume fraction $\Phi$ is large; an example is a living cell with $\Phi$=40\% \citep{Zhou2008}. At finite dilutions there are hydrodynamic corrections to diffusion \citep{Batchelor1976} even for a simple colloidal system of spherical particles. In the literature, crowding has long been treated as an excluded volume interaction at high volume fractions. It is now being realized that electrostatic and hydrodynamic interactions sensitively affect macromolecular dynamics~\citep{Zhou2008,Zhou2009}. As a result, crowding-related effects can be important even at relatively low volume fractions. For example, for a micelle of radius 2 nm in a solution with Debye length $\kappa^{-1} = 1$ nm, the effective radius is 3 nm and $\Phi$=10\% corresponds to $\Phi_{\mathrm{eff}} \approx 34\%$, which already represents a relatively dense colloidal regime. Thus, we generalize macromolecular crowding to refer to all concentrations where excluded volume, electrostatic or hydrodynamic interactions are at play.

The nature of the association of peptides with anionic SDS micelles depends on the details of the electrostatic environment; for example, cationic peptides bind more strongly than their zwitterionic counterparts \citep{Begotka2006}. NMR diffusometry studies have found that peptide binding with anionic SDS micelles and zwitterionic dodecylphosphocholine (DPC) micelles are different, also due to the difference in electrostatic environment \citep{TracyWhitehead2004}. Similarly, it was found that a cell-penetrating peptide (CPP) alters the dynamics and size of neutral and negatively charged bicelles in different ways \citep{Andersson2004}.

PFG-NMR studies have shown that the hydrophobic interaction can play a significant role on the binding of peptides and tripeptides to micelles \citep{Deaton2001,Orfi1998}, as well as neuropeptides to a membrane-mimic environment \citep{ChiradipChatterjee2004}. NMR studies were also carried out to explore the binding of a neuropeptide to SDS micelles in the presence of zwitterionic 3-[(3-cholamidopropyl) dimethylammonio]-1-propanesulfonate (CHAPS) surfactant as a crude model for cholesterol in the biological membrane. These studies showed that having comicelles composed of SDS and CHAPS surfactants inhibits the hydrophobic interaction of the neuropeptide with the core of comicelles \citep{TracyWhitehead2001}. 

Since AMPs are subjects of much interest and also represent an even larger class of amphipathic, helical peptides, the peptide, GAD-2 with a 19-amino acid sequence (FLHHIVGLIHHGLSLFGDR), was selected 
for this study. GAD-2 and a related peptide, GAD-1 with a 21-amino acid sequence, have been identified in recent efforts to discover new AMPs \citep{Fernandes2010,Rise2011,Ruangsri2012}. GAD-2 has recently been shown by NMR and circular dichroism to take on a helical structure in SDS micelles at $40\degree$ C, although it loses a certain amount of its helicity at room temperature (unpublished data). While the GAD-2 -SDS peptide-micelle system chosen is relevant and of current interest in biochemical studies, the goal of this study was to provide a realistic picture of complex formation in peptide-micelle systems in general.

In this work, we used NMR diffusometry to study the interaction between the cationic GAD-2 AMP and an anionic SDS micelle as a membrane mimic environment. In order to do so, we use a simple mathematical model that is utilized to signal the changes in the nature of the macromolecular complexes in a system of nonionic polymer-anionic surfactant system in aqueous solution \citep{SulimanPolymer}. Similar models, based on fast exchange between two or more sites, have been employed previously in surfactant \citep{Stilbs1982, Stilbs1983} and peptide-surfactant systems \citep{AidiChen1995, Deaton2001} and utilized in the latter to extract peptide-micelle binding characteristics. We compare the nature of the resulting peptide-SDS complex with those that form with two much smaller peptides, and are able to identify important distinguishing characteristics. We find, reassuringly, that the most minimal model to extract hydrodynamic size works well for peptides, at least for those with the size (19 residues) of GAD-2; however, one must be careful to avoid the onset of crowding in order to reliably use these simple models.


\section{Materials and Methods}
\label{sec:1}
GAD-2 peptide with average molecular mass $\mathrm{M_{w}}$=2168~g/mole was synthesized using solid phase chemical synthesis employing O-fluorenylmethoxycar$-$ bonyl (Fmoc) chemistry, on a CS336X peptide synthesizer (C S Bio Company, Menlo Park, CA) following the manufacturer's instructions. The peptides were synthesized at a 0.2 mmol scale with a single coupling, using prederivatized Rink amide resin. Resin and all Fmoc amino acids were purchased from C S Bio Company Organic solvents and other reagents used for the synthesis and purification were high- performance liquid chromatography (HPLC) grade or better and purchased from Fisher Scientific (Ottawa, ON) and Sigma-Aldrich Canada (St. Louis, MO). Deprotection and cleavage of the peptides from the resin were conducted with a trifluoroacetic acid (TFA)/water (95:5 by volume) cleavage cocktail followed by cold precipitation with tert-butyl ether. The  crude products were purified by preparative reverse-phase HPLC in a Vydac C-8 column by use of a water/acetonitrile linear gradient with 0.1$\%$ TFA as the ion pairing agent. The molecular weights of the peptides were confirmed by matrix-assisted laser desorption ionization time of flight (MALDI-TOF) mass spectrometry. The purified peptides were lyophilized and stored at 4 $^\circ$C.  

Ala-Gly peptide with $\mathrm{M_{w}}$=146.14 g/mole, Tyr-Leu peptide with $\mathrm{M_{w}}$=294.35 g/mole, and SDS ($99\%$ purity) with $\mathrm{M_{w}}$=288.38 g/mole were purchased from Sigma-Aldrich Canada (St. Louis, MO) and were used as received without further purification. Deuterium oxide $\mathrm{D_{2}O}$ with $99.9\%$ isotopic purity was purchased from Cambridge Isotope Laboratories (St. Leonard, Quebec). 

\begin{table*}[ht]
\centering
\caption{Sample nomenclature. All samples were made with $\mathrm{D_{2}O}$ as a solvent, and unless stated have 0.1 M sodium oxalate buffer in them. Final concentrations [SDS] were achieved by mixing different stock solutions. The molar ratio $R = $[peptide]$/$[SDS]=30 was kept constant for GAD-2 solutions.}
\label{Table1}
\begin{tabular}{ |l|l|l|l|l|l|l|}\hline
Abbreviation &Final [SDS]\\
\hline
SDS-buf & 2-187 mM  \\
GAD-2--SDS &1-80 mM \\ 
      
Ala-Gly--SDS &2-60 mM \\     
Tyr-Leu--SDS &2-60 mM \\     
\hline    
\end{tabular}
\end{table*}


GAD-2--SDS, Ala-Gly--SDS, Tyr-Leu--SDS, and SDS samples were prepared with compositions according to table~\ref{Table1}. The molar ratio (R) of SDS concentration to peptide concentration in GAD-2--SDS samples was held constant ($\mathrm{R=[SDS]/}$[GAD-2]=30). The concentration of dipeptides (Ala-Gly and Tyr-Leu) in Ala-Gly--SDS and Tyr-Leu--SDS systems was 2 mM. The pH value for all samples was adjusted to be 4 by the addition of sodium deuteroxide or deuterium chloride. All samples were made with $\mathrm{D_{2}O}$ as solvent and, unless otherwise stated, have 0.1 M sodium oxalate buffer ($\mathrm{Na_{2}C_{2}O_{4}}$) in them. Sodium oxalate buffer was used in previous NMR studies to adjust the pH of SDS micelle-peptide solutions~\citep{Orfi1998,Deaton2001}. It is effective as a buffer for pH below 5, where the histidine-rich GAD-2 peptide is expected to have a net positive charge. Moreover, the chemical structure of sodium oxalate does not include protons in it. As a result, the one dimensional proton NMR spectra do not include buffer peaks that might overlap with SDS and peptides peaks.

The self-diffusion measurements were carried out in a diffusion probe (Diff30) and with maximum field gradient 1800 G/cm (applied along the z-axis) at a resonance frequency of 600 MHz on a Bruker Avance II spectrometer. The maximum gradient used in this work was 300 G/cm. Diffusion was measured with a pulsed-field gradient stimulated-echo sequence \citep{s1} with (almost square) trapezoidal gradient pulses. The diffusion coefficient of a molecule in aqueous solution is obtained from the attenuation of the signal according to the equation \citep{s1}

\begin{equation}
\label{SignalAttenuation}
\mathrm{\ln\left(\frac{S(k)}{S(0)} \right)=-Dk}
\end{equation}
where S(k) is the \textgravedbl intensity\textacutedbl of the signal (the integration of the relevant peak region) in the presence of field gradient pulse, S(0) is the intensity of the signal in the absence of
field gradient pulse, $\mathrm{k=(\gammaup\deltaup g)^{2}(\Delta-\deltaup/3)}$ is a generalized gradient
strength parameter, $\mathrm{\gammaup=\gammaup^{H}=2.6571\times 10^{8}~T^{-1}s^{-1}}$ is the gyromagnetic ratio of the $\mathrm{^{1}H}$ nucleus, $\mathrm{\deltaup=2~ms}$ is the duration of the field gradient pulse, $\mathrm{\Delta=100~ms}$ is the time period between the two field gradient pulses, and g is the amplitude of the field gradient pulse.

\section{Results and Discussion}

\begin{figure}[htp]
\centering
\subfloat[]{%
\label{SDS.png}\includegraphics[scale=0.16]{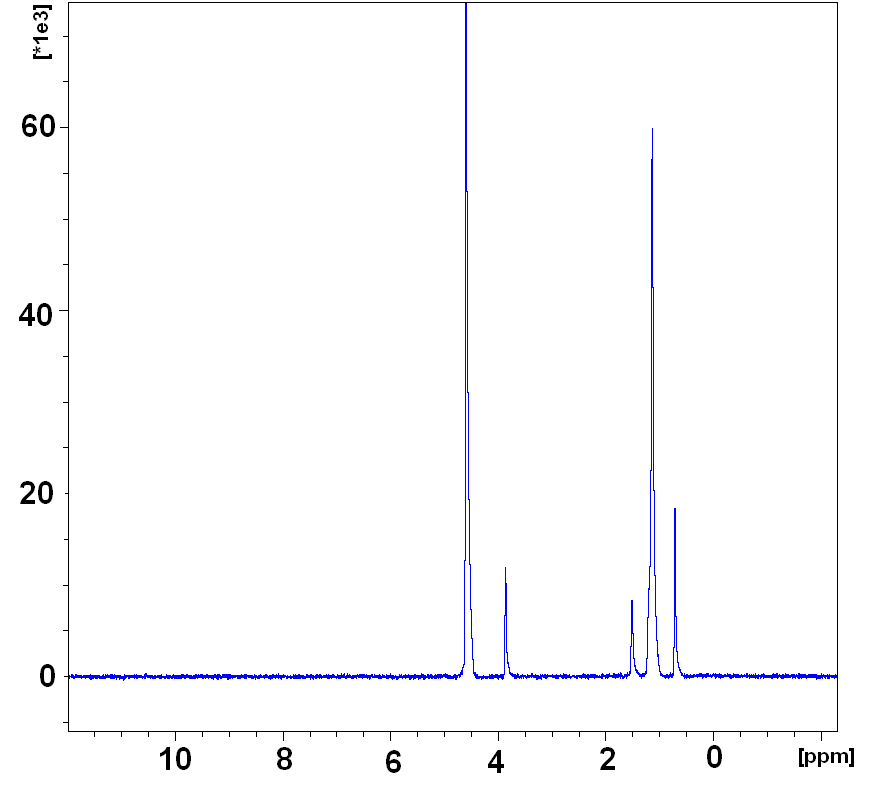}}
\hspace{1in}
\subfloat[]{%
\label{Gad2SDS.png}\includegraphics[scale=0.25]{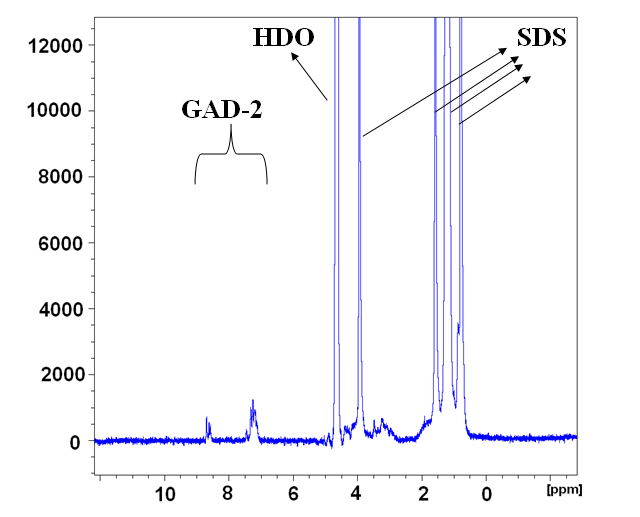}}
\caption{1D $\mathrm{^{1}H}$ NMR spectrum for (a) a peptide-free SDS sample with $\mathrm{[SDS]=6~mM}$ (b) a GAD-2--SDS sample with [SDS]=60 mM and [GAD-2]=2 mM. 
Sample temperature is 298 K.}
\label{OneDimensionalSpectra}
\end{figure}

Complementary NMR-based techniques were utilized in order to identify components for different samples based on their one-dimensional NMR spectra and to extract parameters such as self-diffusion coefficients. The one-dimensional (1D) proton NMR spectra at a resonance frequency of 600 MHz on a Bruker Avance II spectrometer and at sample temperature 298 K are shown in figure~\ref{OneDimensionalSpectra}. In all cases the trace signal of HDO in $\mathrm{D_{2}O}$ is the most dominant peak (at $\approx$ 4.7 ppm); however the HDO, peptide and SDS peaks are all spectrally separable. In NMR, chemical shifts can be utilized to provide informatoin about the structure and the change in the chemical environment of molecules. For example, it was found \citep{Larive2005} that both the chemical shift and the observed diffusion coefficient are affected by complexation. However, in our work, we specifically prepared our samples so that the SDS concentration was varied, but with the molar ratio R = [SDS]/[GAD-2] held constant. We thus do not see a change in either linewidths or chemical shifts as a function of SDS concentration.


\begin{figure}[htp]
\centering
\subfloat[]{%
\label{SDSSignalAttenuation.eps}\includegraphics[scale=0.30]{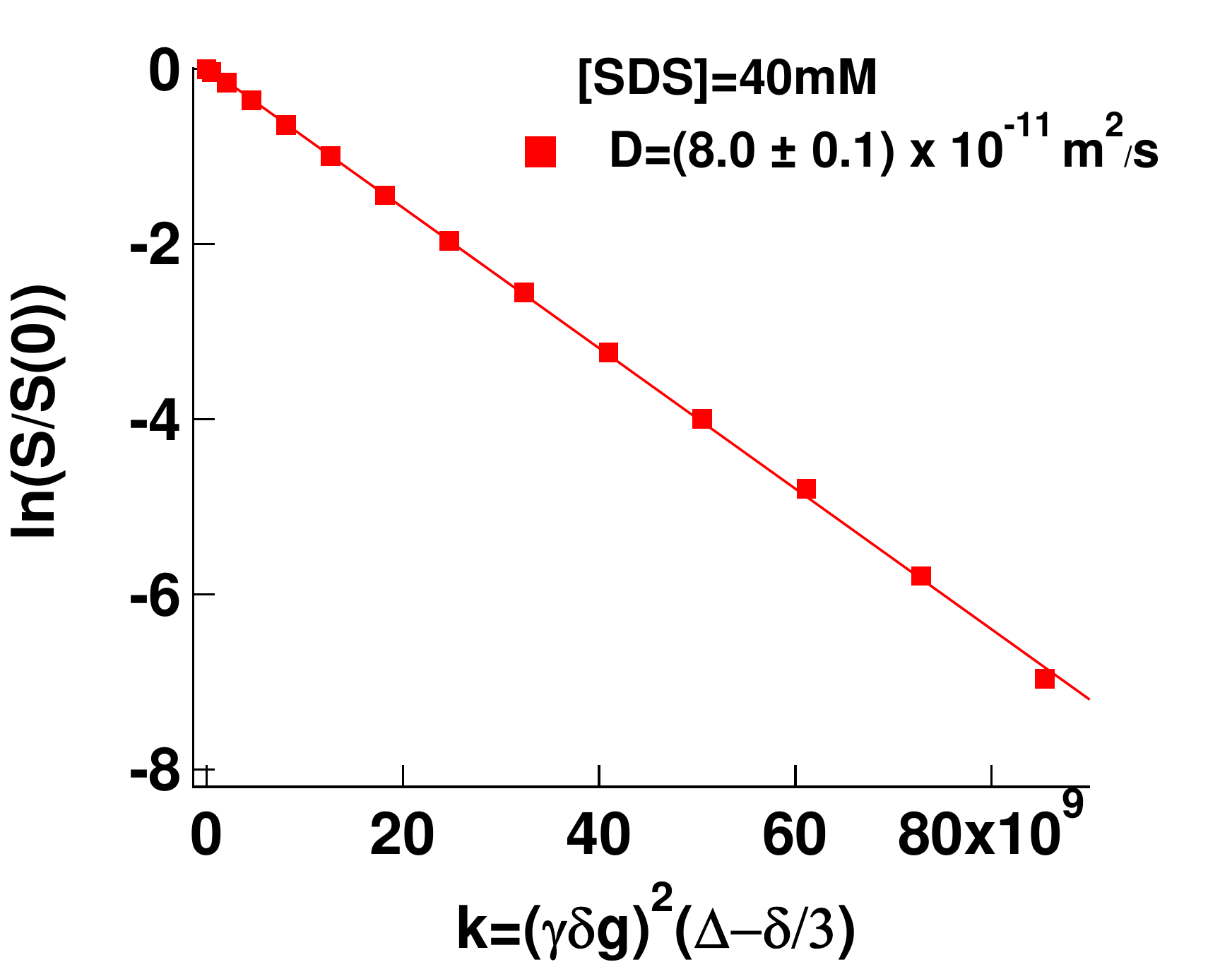}}
\hspace{1cm}
\subfloat[]{%
\label{Gad2SignalAttenuation.eps}\includegraphics[scale=0.30]{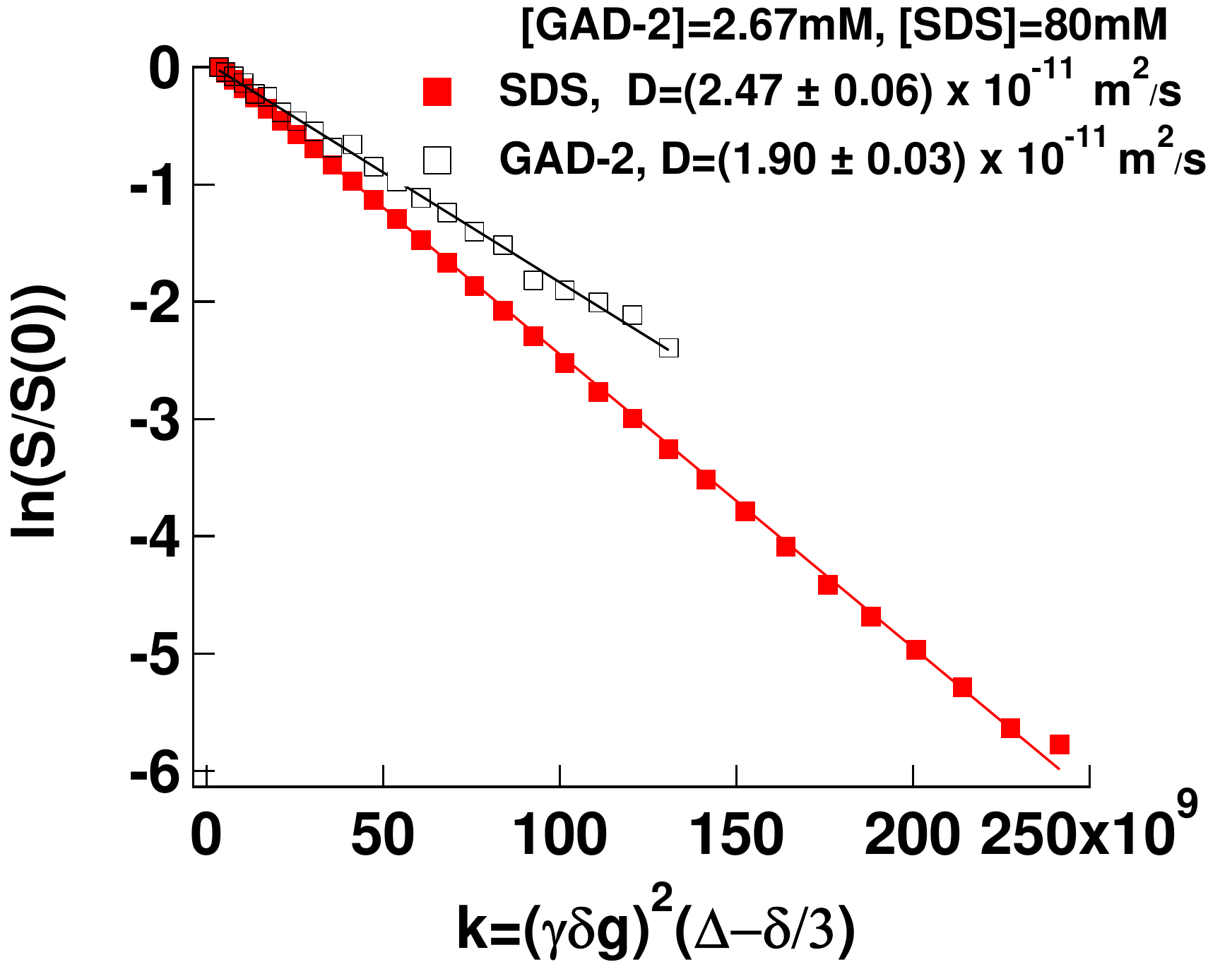}}
\caption{The attenuation of the signal S(k)/S(0) on a log scale versus $\mathrm{k=(\gammaup\deltaup g)^{2}(\Delta-\deltaup/3)}$ for (a) a peptide-free SDS sample with $\mathrm{[SDS]=40~mM}$ and 0.1 M sodium oxalate buffer (b) a GAD-2--SDS sample with [SDS]=80 mM, [GAD-2]=2.67 mM, and 0.1 M sodium oxalate buffer. $\mathrm{\deltaup=2~ms}$ and $\mathrm{\Delta=100~ms}$.The errors in the values of the diffusion coefficients represent the uncertainty in the slope obtained from a linear fit to $\ln(S/S_0)$ $vs$ $k$.Typical values of $\mathrm{R^{2}}$ are of order 0.998.}
\label{SignalAttenuationCurves}
\end{figure}

In this work, we carried out experiments with peptide at varying SDS concentrations in the presence of sodium oxalate buffer. We also performed experiments on pure SDS solutions as well as buffered SDS solutions for comparison. Figure~\ref{SignalAttenuationCurves} shows the signal attenuation and the self-diffusion coefficients for SDS and peptides in a buffered peptide-free SDS sample and GAD-2--SDS sample. The signal attenuation in all samples was observed to be monoexponential. 

This suggests that the exchange of SDS molecules between the SDS in micelles and in free solution must be very rapid in the NMR time scale. The values of the observed diffusion coefficients were calculated from the monoexponential decays using equation~\ref{SignalAttenuation}. For peptide-free SDS solutions prepared with sodium oxalate buffer (figure~\ref{SDSSignalAttenuation.eps}), the signal attenuation of SDS was obtained by integrating the area under the spectral region between 0 to 4 ppm. For the GAD-2--SDS system, the spectral ranges from 0 to 4 ppm and 7 to 9 ppm were used to obtain SDS and GAD-2 signal attenuation, respectively. In each case the SDS and peptide spectral regions were chosen to ensure a clear spectral separation. 

\subsection{Diffusometry}
\subsubsection{Surfactant Solutions and Analysis Methods}

\begin{figure*}[htp]
\centering
\subfloat[]{%
\label{SDSdiffusionCurve1}\includegraphics[scale=0.3]{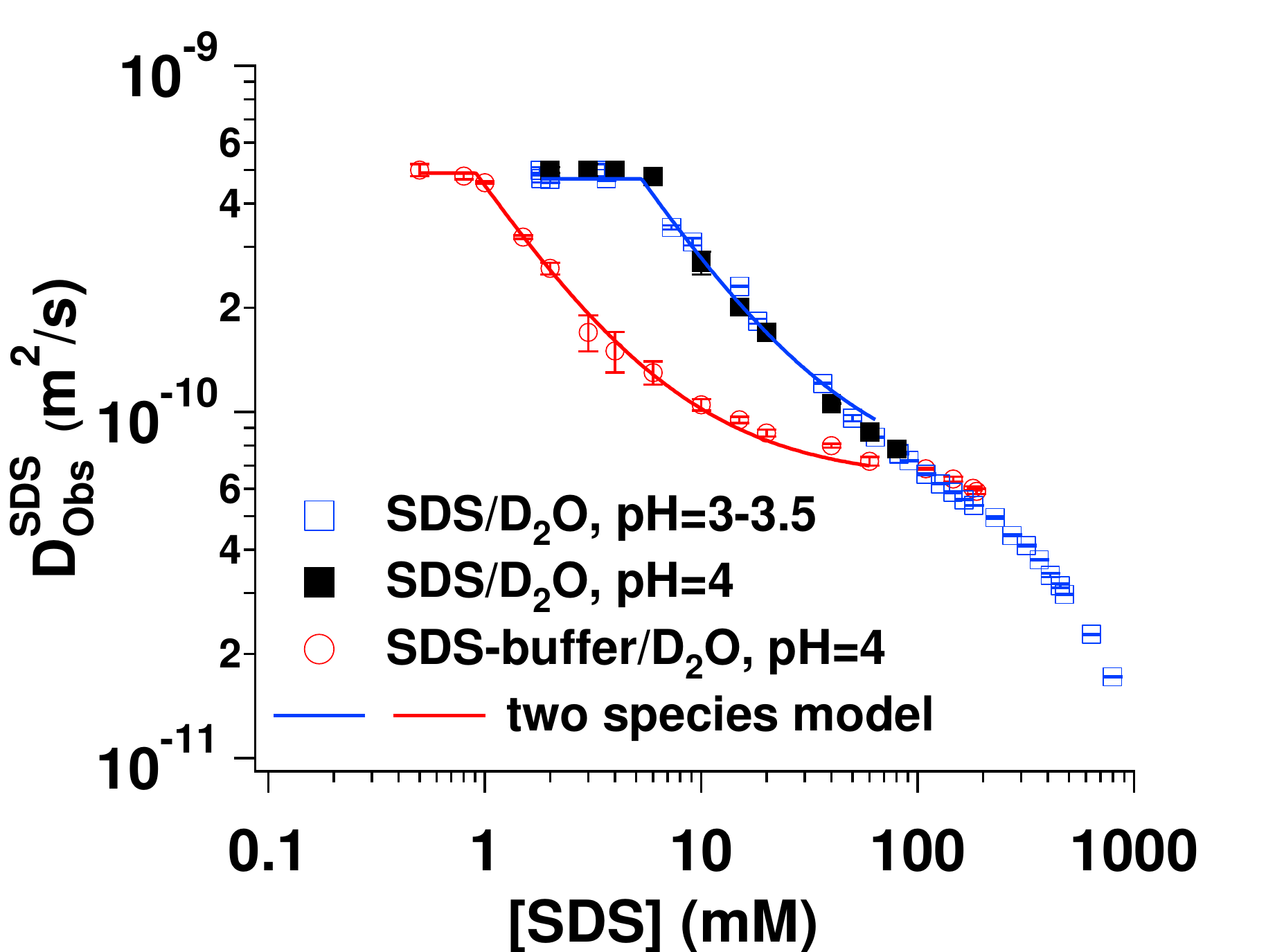}}
\hspace{1cm}
\subfloat[]{%
\label{FreeSDSFractionAndSDSFreeConcentrationSDS.eps}\includegraphics[scale=0.3]{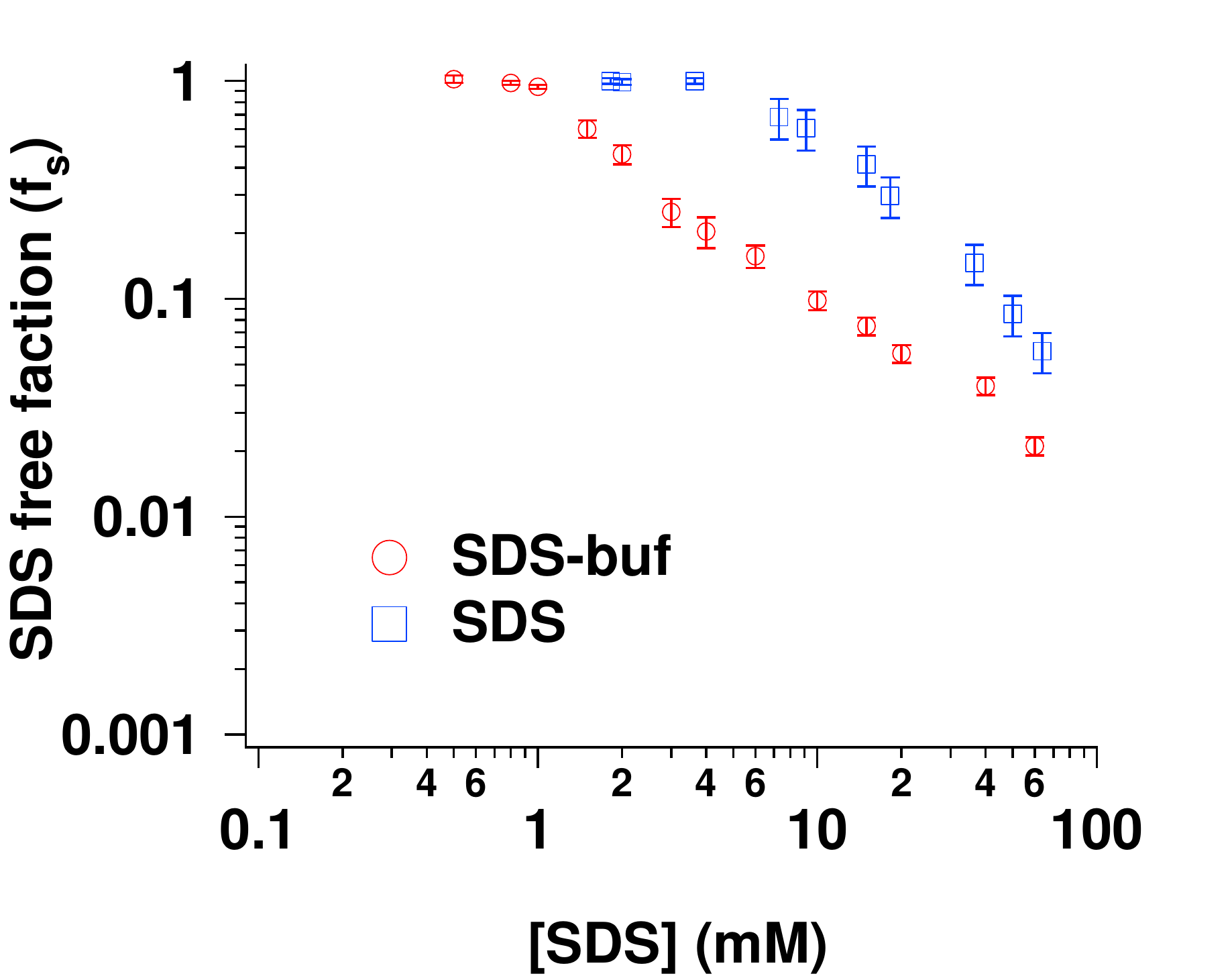}}
\caption{Self-diffusion coefficient in peptide-free SDS solutions. (a) $\mathrm{D}$ versus SDS concentration [SDS] for solutions with sodium oxalate buffer (pH=4) (red open circles), and unbuffered, with pH=3-3.5 (blue open squares), and with pH=4 (black filled squares). (b) Fraction ($\mathrm{f_s}$) of free SDS with and without sodium oxalate buffer.}
\label{SDS1}
\end{figure*}

Figure~\ref{SDSdiffusionCurve1} shows the self-diffusion coefficient of SDS in 3 peptide-free SDS systems: one with sodium oxalate buffer with pH=4 (red open circles), and two without sodium oxalate buffer. Of the unbuffered solutions one was with pH unadjusted but measured to be between 3 and 3.5 (blue open squares), and one with the pH=4 (black filled squares). Below [SDS] = 60 mM, the SDS diffusion coefficient $\mathrm{D^{SDS}_{Obs}}$ for unbuffered solutions at different pH are indistinguishable
from each other,
 while values in the buffered solution are much lower.

The pulsed-field-gradient signal attenuation is monoexponential for all samples (figure~\ref{SignalAttenuationCurves}). This implies that the exchange of SDS molecules between the SDS in micelles and in free solution is rapid in the NMR time scale. Previous studies \citep{Soderman1994,Stilbs1982,Stilbs1983} showed that a surfactant molecule visits more than one environment over very short timescales, and  interpreted the observed diffusion coefficients using a two-site exchange model. In all the systems considered here, the SDS surfactant can either be a free monomer in solution or associated with a surfactant-rich aggregate. 
The observed self-diffusion coefficient of SDS is thus a linear combination of the self-diffusion coefficient $\mathrm{D^{SDS}_{free}}$ of the free molecules in bulk and that of the bound molecules in the micelle $\mathrm{D^{SDS}_{micelle}}$ in peptide-free solutions or a peptide-SDS complex $\mathrm{D^{SDS}_{Aggr}}$

\begin{align}
\label{Approach11}
& \mathrm{D_{Obs}^{SDS}=D^{SDS}_{free}},\, &\, \mathrm{[SDS]\leq C_0}, \nonumber\\ 
& \mathrm{D_{Obs}^{SDS}= \left(D^{SDS}_{free}-D^{SDS}_{Aggr}\right) (f_s)+ D^{SDS}_{Aggr}}, &\, \mathrm{[SDS]>C_0}
\end{align}
where $\mathrm{f_s}= \mathrm{[SDS]_{free}/[SDS]}$ is the fraction of free SDS molecules, $\mathrm{D^{SDS}_{Aggr}}$ is either the micellar diffusion coefficient in peptide-free samples, or the diffusion coefficient of the micelle-peptide complex, and $\mathrm{C_0}$ refers to the critical (micellar or aggregation) concentration (CMC or CAC), and $\mathrm{[SDS]}$ is the total SDS concentration. A key assumption of the model is that there are only two distinct species, the free and the aggregate states; however, as will be seen
later, we are able to check for self-consistency of the model.


For simple spherical micelle systems, buffered and unbuffered peptide-free SDS solutions, $\mathrm{[SDS]_{free}=CMC}$ for $\mathrm{[SDS]>CMC}$. Therefore, equation \ref{Approach11} has 3 parameters, $\mathrm{C_0=CMC}$, $\mathrm{D^{SDS}_{free}}$ and $\mathrm{D^{SDS}_{micelle}}$. Fitting the buffered peptide-free SDS solution to the two-species model in equation~\ref{Approach11} yields the parameters  $\mathrm{D^{SDS}_{free}=(4.90\pm 0.07)\times 10^{-10}~m^{2}/s}$,  $\mathrm{D^{SDS}_{micelle}=(6.3\pm 0.4)}$ $\mathrm{\times 10^{-11}~m^{2}/s}$, and $\mathrm{CMC=(0.91\pm 0.02)}$ mM, while for the unbuffered peptide-free SDS solution $\mathrm{D^{SDS}_{free}=(4.71\pm 0.08)\times 10^{-10}~m^{2}/s}$,  $\mathrm{D^{SDS}_{micelle}=(6.1\pm 0.9)\times 10^{-11}}$ $\mathrm{m^{2}/s}$, and $\mathrm{CMC=(5.3\pm 0.2)~mM}$.    

The main physical insight hidden in these curves is the onset of crowding. While the unbuffered and buffered solutions have very different dynamics at low [SDS], they both report a constant and similar micelle size upto 60 mM. Above 60 mM, the observed diffusion is reporting on micellar diffusion in an environment where inter-micellar interactions cannot be neglected. Two effects are thus inseparable in either dynamic light scattering or pulsed-field-gradient NMR: reduction in micellar diffusion coefficient due to increase in hydrodynamic size, and increase in hydrodynamic interactions between complexes. Such an effect of hydrodynamic interactions has indeed been previously noted~\citep{TadashiAndo2010}. 


\subsubsection{Peptide: GAD-2}
When the size of a hydrophobic peptide is large enough that surfactant motion is rapid on the timescale of peptide motion, the peptide is expected to be associated with several surfactant molecules and there should never be free peptide, i.e. the peptide binding fraction is close to 1. For example, in the GAD-2--SDS system, since the concentration of SDS is 30 times higher than GAD-2 concentration ($\mathrm{R=[SDS]/[GAD-2]=30}$), we know that there is unlikely to be free peptide: we will test this assumption soon.

In this case, $\mathrm{D^{SDS}_{Aggr}}$=$\mathrm{D^{Peptide}_{Aggr}} \approx \mathrm{D^{Peptide}}$.
Using this additional information allows us to use the two-site model even if the $\mathrm{D^{SDS}_{Obs}}$ versus $\mathrm{1/[SDS]}$ relationship is not linear. The only proviso is that the overall particulate volume fraction must always be small enough that hydrodynamic effects are negligible. For the peptide-free SDS system, we have seen that this is generally true for concentrations below 60 mM, or volume fractions below 0.04. For GAD-2--SDS system, the size of an GAD-2--SDS aggregate is expected to change with SDS concentration. Therefore, the concentration $\mathrm{[SDS]_{free}}$ of free SDS monomers is expected to change in the SDS concentration regime above CAC. We may simply rewrite and rearrange equation~\ref{Approach11} for $\mathrm{[SDS]>C_{0}}$ but with $\mathrm{D^{SDS}_{Aggr}}$=$\mathrm{D^{Peptide}}$,  
\begin{equation}
\label{SDSfreeFraction1}
\mathrm{f_s([SDS])=\mathrm{\frac{[SDS]_{free}}{[SDS]}}=\frac{D^{SDS}_{Obs}-D^{Peptide}}{D^{SDS}_{free}-D^{Peptide}}}.
\end{equation}

\begin{figure*}[htp]
\centering
\subfloat[]{%
\label{Gad2diffusionCurve}\includegraphics[scale=0.3]{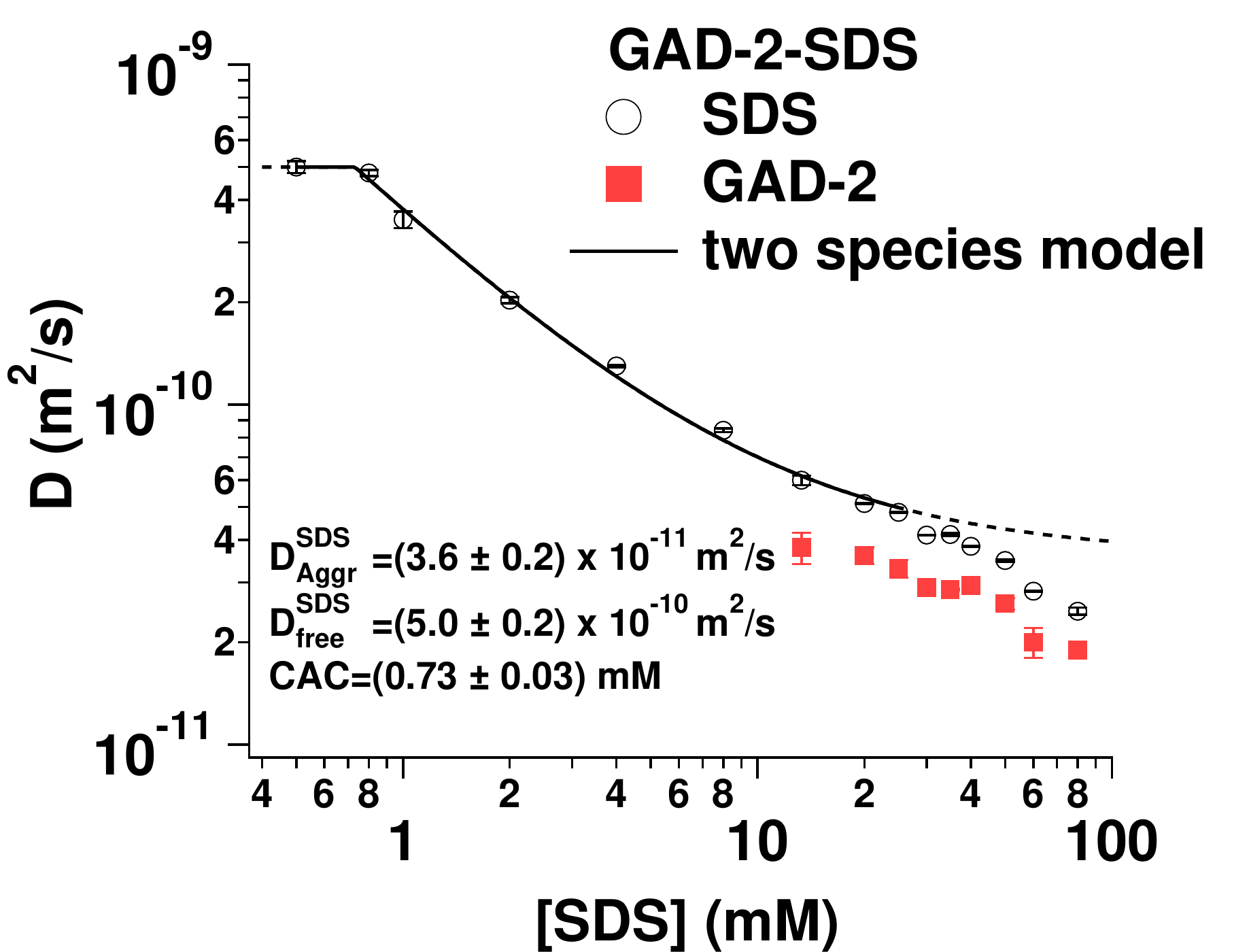}}
\hspace{0in}
\subfloat[]{%
\label{FreeSDSFractionAndSDSFreeConcentrationGad2.eps}\includegraphics[scale=0.3]{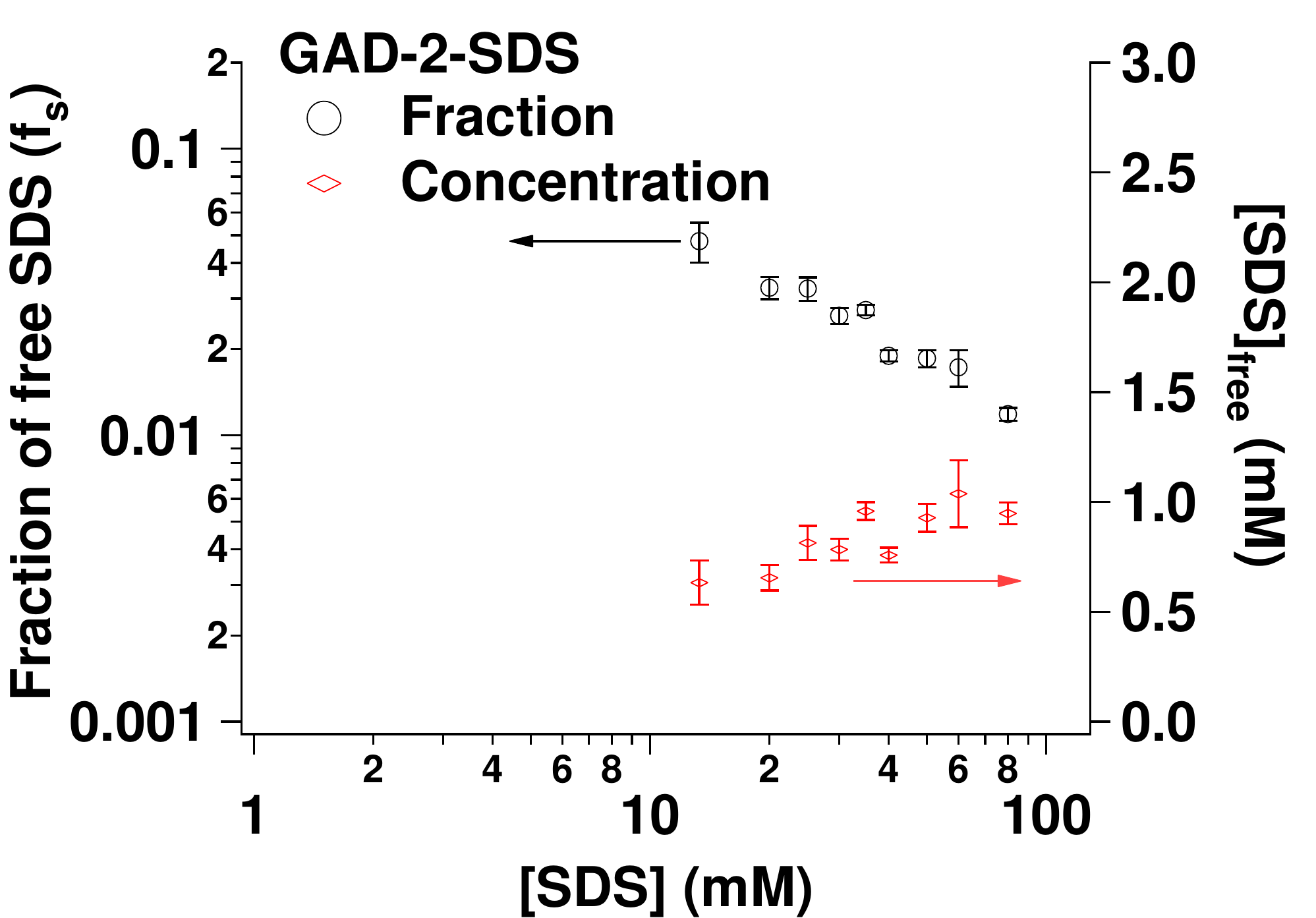}}
\hspace{0in}
\caption{(a) Self-diffusion coefficient of GAD-2 and SDS in a GAD-2--SDS system with R=[SDS]$/$[GAD-2]=30 versus SDS concentration [SDS] (b) Fraction ($\mathrm{f_{s}}$) of free SDS and concentration ($\mathrm{[SDS]_{free}}$) of free SDS versus SDS concentration $\mathrm{[SDS]}$}.
\label{Gad2A}
\end{figure*}

Figure~\ref{Gad2diffusionCurve} shows the self-diffusion coefficient of GAD-2 and SDS in the GAD-2--SDS system. We measured the self-diffusion of GAD-2 in the SDS concentration range that is higher than 13.3 mM. Due to experimental limitations (small value of signal-to-noise ratio), we were not able to extract the self-diffusion coefficient of GAD-2 in the SDS concentration range below 13.3 mM, but we were able to measure the surfactant diffusion. 

The SDS self diffusion coefficient is fit well to the two species model for [SDS]$\leq$25 mM (figure~\ref{Gad2diffusionCurve}, solid line), and it deviates from the fit for higher SDS concentration (figure~\ref{Gad2diffusionCurve}, dotted line). 
The resulting fit parameters are $\mathrm{D^{SDS}_{free}}$ = $\mathrm{(5.0\pm 0.2)\times 10^{-10}~m^{2}/s}$, $\mathrm{D^{SDS}_{Aggr}}$ $=\mathrm{(3.6\pm 0.2)\times 10^{-11}~m^{2}/s}$, and CAC= (0.73$\pm$ 0.03) mM. We now test the assumption that there is no free peptide. Using a two-site exchange model similar to Equation 2, but for the peptide (with $\mathrm{D^{Peptide}_{Obs}=3.8 \times 10^{-11}~m^{2}/s}$ at [SDS]=13 mM and $\mathrm{D^{Peptide}_{free} \ge 1.6 \times 10^{-10}~m^{2}/s}$, the value in SDS-free buffered peptide system at [GAD-2]=2 mM, and $\mathrm{D^{Peptide}_{Aggr}}=\mathrm{D^{SDS}_{Aggr}}$), we calculated the fraction of free peptide at [SDS]=13 mM to be $\le$ 1.6$\%$. Previous studies~\citep{Gao1998} reported the partitioning of adrenocorticotropic hormone (ACTH) peptides in SDS and DPC micelles. There too, the fraction of ACTH bound to SDS is over 99$\%$.

PFG-NMR can be used to get spectrally-resolved diffusion coefficients~\citep{Johnson1992,Johnson1993,Johnson1994,DWu1994,Byrd1995}. 
Using both SDS and peptide diffusion coefficients as a function of [SDS], we extract the fraction ($\mathrm{f_{s}}$) of free surfactant in the monomer state in the aqueous solution as well as the concentration of free surfactant $\mathrm{[SDS]_{free}}$. This is shown in figure \ref{FreeSDSFractionAndSDSFreeConcentrationGad2.eps}. 
With increasing surfactant concentration, $\mathrm{f_{s}}$ decreases while $\mathrm{[SDS]_{free}}$ rises from 0.7 mM (close to the CAC) to $\approx$ 1 mM (close to the CMC). This is consistent with the picture \citep{SulimanPolymer,jones} that the concentration of free surfactant above the CAC/CMC is equal to the value of the CAC/CMC. In the peptide-SDS system, and similar to the behavior in the nonionic polymer--anionic surfactant  (poly(ethylene)oxide--SDS) system \citep{SulimanPolymer}, the free concentration rises further until it reaches the CMC value in the buffered solution.

\begin{figure}[htbp]
\centering
\subfloat[]{%
\label{Gad2SDSApparentRadiusGyration.eps}\includegraphics[scale=0.3]{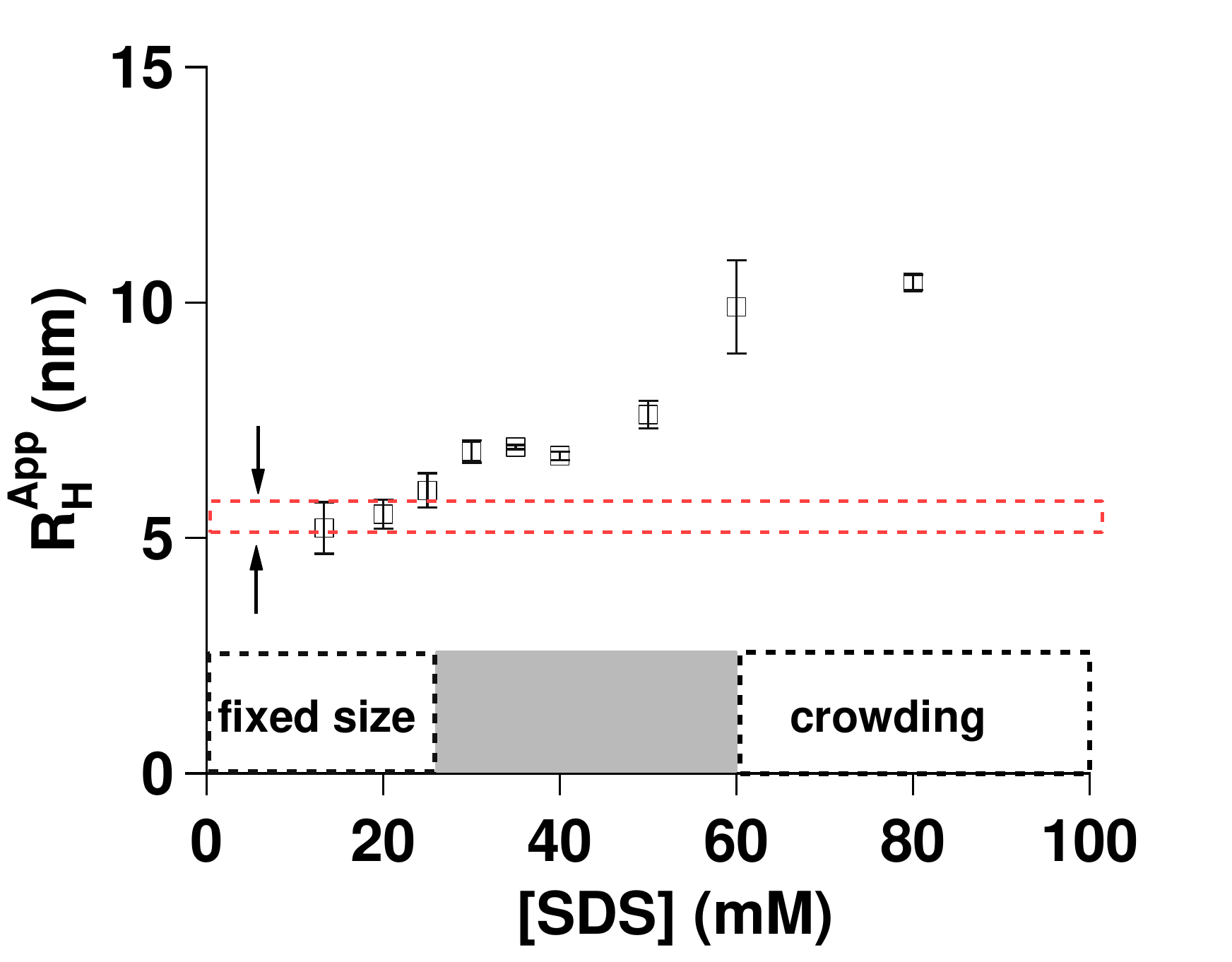}}
\hspace{0cm}
\subfloat[]{%
\label{SDSAggregationGad2.eps}\includegraphics[scale=0.3]{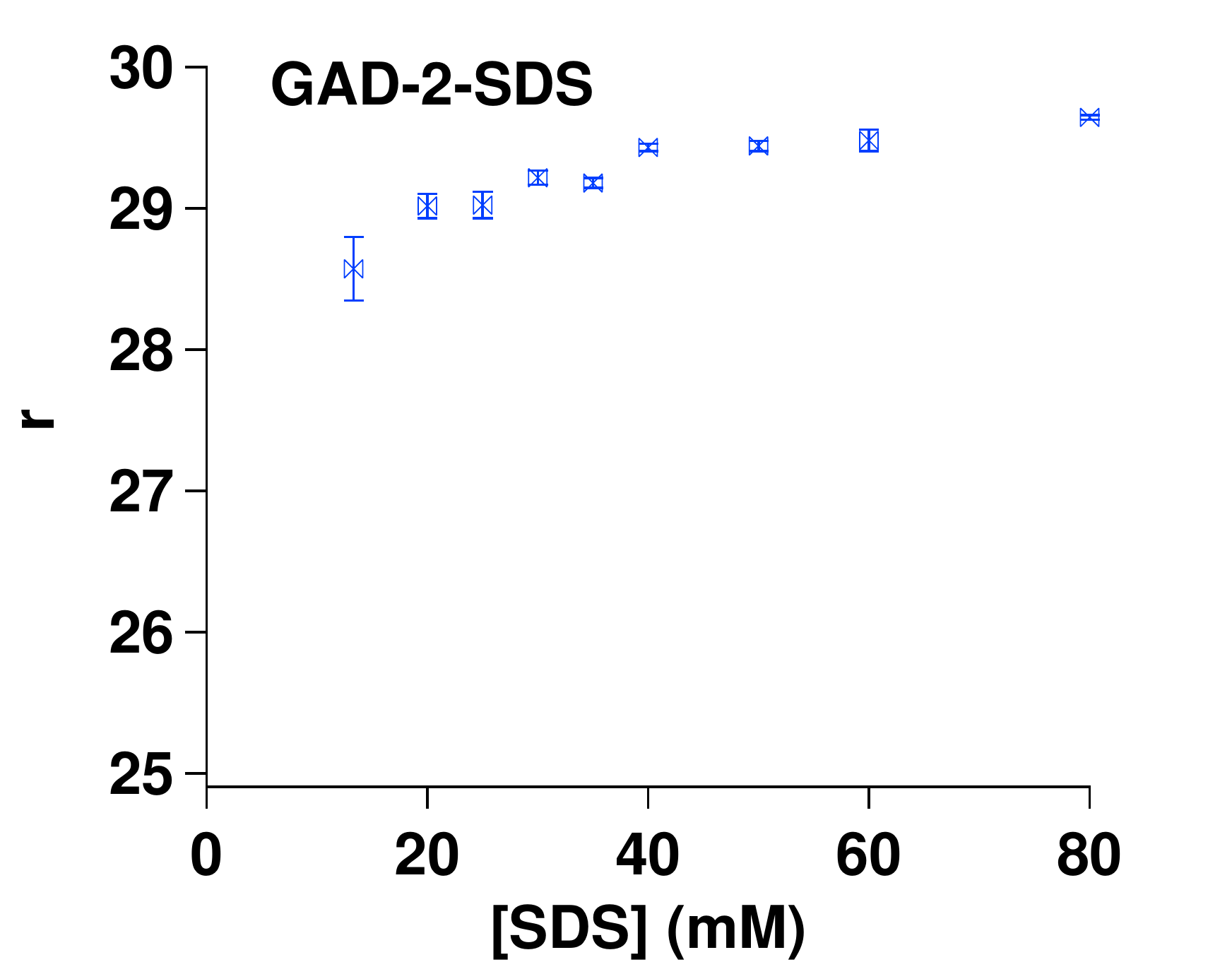}}
\caption{(a) The apparent hydrodynamic radius ($\mathrm{R^{App}_{H}}$), extracted from the peptide diffusion coefficient, of GAD-2--SDS complexes versus SDS concentration [SDS]. The horizontal dashed line is the value of apparent hydrodynamic radius ($\mathrm{R^{App}_{H}}$ =5.5 $\pm$ 0.3 nm) obtained $via$ the SDS aggregate diffusion coefficient ($\mathrm{D^{SDS}_{Aggr}=(3.6 \pm 0.2) \times 10^{-11}~ m^{2}/s}$ ) at low SDS concentration [SDS]$\le$ 25 mM, where the two species model is valid. This value is in agreement with the peptide diffusion coefficient. At high SDS concentration [SDS] $\geq$ 60 mM. The diffusion coefficient measured gives no information about the true hydrodynamic radius. The intermediate SDS concentration regime, denoted by the gray area, is the regime in which either complex size is indeed increasing with concentration or hydrodynamic interactions between complexes is slowing down the motions. (b) The ratio of SDS molecules to peptide molecules in a complex (r) versus SDS concentration ($\mathrm{[SDS]}$) for GAD-2--SDS samples.}
\label{Gad2B}
\end{figure}

Next, we estimate the hydrodynamic radius $\mathrm{R_{H}}$ of GAD-2--SDS complexes (figure~\ref{Gad2SDSApparentRadiusGyration.eps}) using the Sutherland~$-$~Stokes~$-$~Einstein equation \citep{jones}
\begin{equation}
\label{RadiusOfGyration}
\mathrm{R_{H}}=\frac{\mathrm{K_{B}T}}{\mathrm{6\pi\etaup\, D}}
\end{equation}
where $\mathrm{K_{B}}$ is the Boltzmann constant, T is the absolute temperature, and $\etaup$ is the solvent viscosity ($\etaup_{D_{2}O}$=1.1 mPa.s). 
The hydrodynamic radius $\mathrm{R_{H}}$ is obtained from the peptide diffusion ($\mathrm{D=D^{Peptide}}$, open squares in figure~\ref{Gad2SDSApparentRadiusGyration.eps}) as well as from the fitted value of $\mathrm{D^{SDS}_{Aggr}}$ obtained from the concentration dependence of the surfactant diffusion (dashed red line in figure~\ref{Gad2SDSApparentRadiusGyration.eps}). 
For [SDS] $<$ 25 mM the hydrodynamic radii obtained via peptide diffusion and surfactant diffusion are roughly the same, with a value of approximately 5.5 nm. Interestingly, $\mathrm{R_{H}}$ (obtained from peptide diffusion $\mathrm{D^{Peptide}}$) increases as a function of SDS concentration to about 10 nm at 60 mM, less than a factor of two increase.

Plotted in figure~\ref{SDSAggregationGad2.eps} is the variation in the ratio of SDS molecules to peptide molecules in a complex $\mathrm{r=}$ $\mathrm{([SDS]-[SDS]_{free})/([SDS]/R)}$ = $(1 - \mathrm{f_{s}})$ R exhibits a very slight increase, from $\approx$~28 to 29, and approaches R = 30 asymptotically. We need to understand how the aggregate size changes in order to accommodate the two-fold increase in the hydrodynamic radius $\mathrm{R_{H}}$; we will return to this point later.

\subsubsection{Comparison with smaller dipeptides}

In order to study the effect of peptide size on the dynamics of peptide-SDS complexes, and to ensure consistency with previous work on small peptides~\citep{Deaton2001}, diffusometry was carried out to quantify complex formation of SDS with the dipeptides Ala-Gly and Tyr-Leu. The measured diffusion coefficients for both the SDS and peptides are consistent with those measured at one SDS concentration in that previous work~\citep{Deaton2001}.

\begin{figure}
\centering
\includegraphics[scale=0.3]{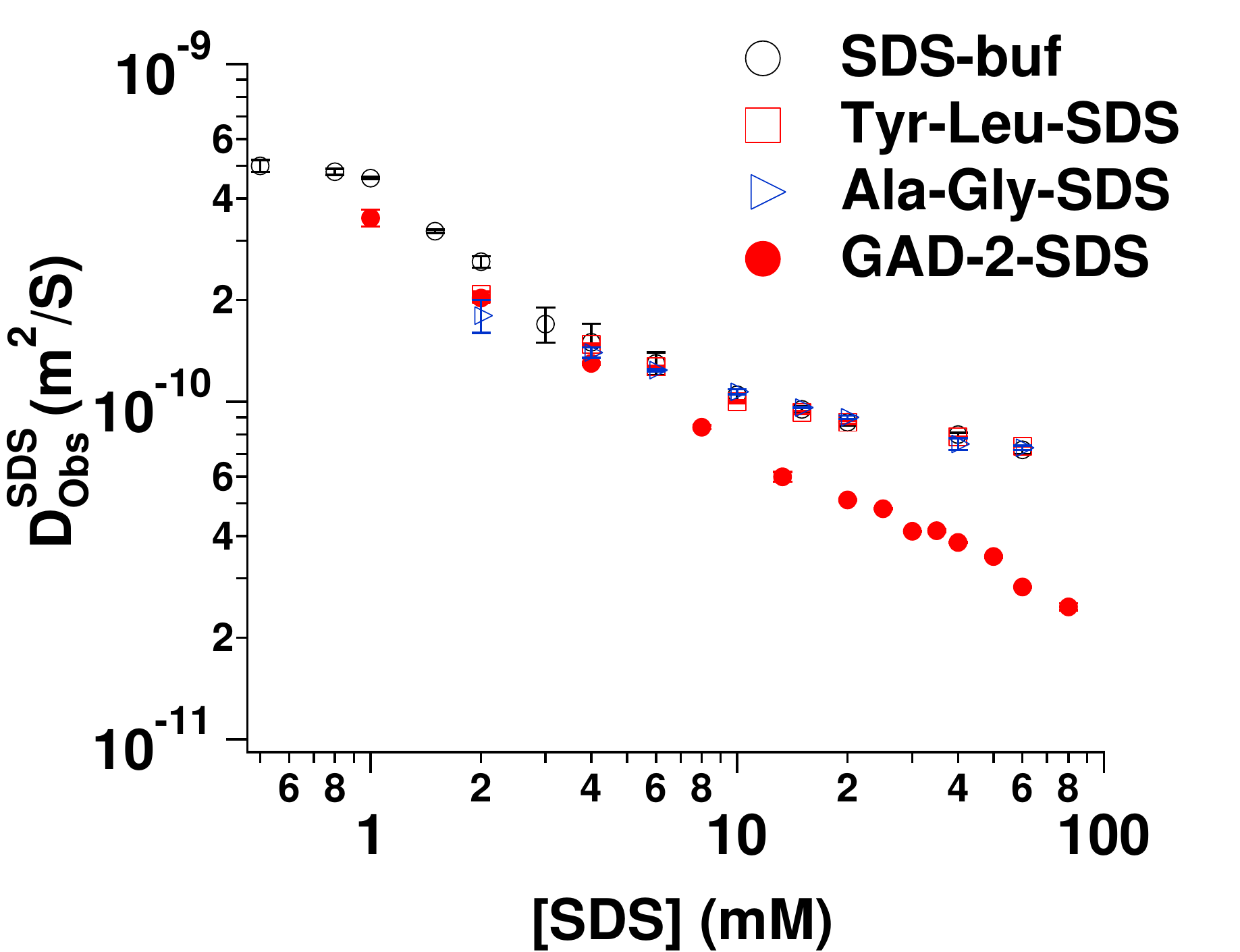}
\caption{\label{SDSDiffusionForDifferentPeptides} Self-diffusion coefficient of SDS in all the systems studied: peptide-free SDS system with sodium oxalate buffer, Tyr-Leu--SDS, Ala-Gly--SDS, and GAD-2--SDS with R=[SDS]$/$[GAD-2]=30 samples.}
\end{figure}

A plot of the SDS self-diffusion coefficient for all systems in the current study in one graph (figure~\ref{SDSDiffusionForDifferentPeptides}) shows clearly that SDS diffusion looks similar for the systems with small di-peptides (Ala-Gly and Tyr-Leu) and the peptide-free SDS system with sodium oxalate buffer. This suggests that the fraction of free SDS in the Tyr-Leu--SDS and Ala-Gly--SDS systems is similar to the fraction of free SDS in the peptide-free SDS system with buffer (figure~\ref{FreeSDSFractionAndSDSFreeConcentrationSDS.eps}). On the other hand, SDS diffusion looks very different for the system with long peptide (GAD-2--SDS system), suggesting that the GAD-2--SDS complexes are very different from the Ala-Gly--SDS and Tyr-Leu--SDS complexes, which are essentially indistinguishable from micellar aggregates with no peptide.  

This means that the peptide-micelle binding characteristics of the Tyr-Leu and Ala-Gly dipeptides are different from the much longer GAD-2 peptide. Also, this indicates that GAD-2 significantly disrupts the micellar aggregate. This conclusion likely extends to other long and hydrophobic peptides.

\section{Conclusion}



NMR-based techniques have been utilized in this work to study the nature of peptide-micelle complexes in a buffered 19-residue antimicrobial peptide (the GAD-2--SDS system). 
First, we examined the impact of the buffer (figure~\ref{SDSdiffusionCurve1}). Varying the pH over a small range in the absence of a buffer shows no effect on the micellar structure, while the CMC is lower in the presence of the buffer. The addition of sodium salts more effectively screens the charge on the micelle. In other work it has been found to result in larger stable micelles~\citep{Haataja2009,Jones1988} and lower critical micellar concentrations~\citep{Krastev2009}. 


For pure (peptide-free) SDS solutions, the observed diffusion coefficients of surfactant SDS molecules for buffered and unbuffered solutions merge at surfactant concentrations [SDS] $>$ 60 mM. In addition, the linear two species model (equation~\ref{Approach11}) is robustly valid below [SDS]=60 mM, with micelle size being independent of SDS concentration. This is similar to the findings in previous work for a system of anionic surfactant (SDS)-nonionic polymer polyethylene oxide (PEO)~\citep{SulimanPolymer} where this concentration was identified as the onset of macromolecular crowding: this refers to the excluded volume effect of one macromolecule with respect to another~\citep{Zhou2008}. Our primary finding is that [SDS]=60 mM signals the concentration beyond which one cannot, even in principle, extract hydrodynamic radii or aggregate ratios.

At low surfactant concentrations ([SDS] $<$  25 mM), the observed diffusion coefficient of SDS (figure~\ref{Gad2diffusionCurve}) is well described by the two-species model in equation~\ref{Approach11}, with both monomer and aggregate having a size that does not depend on SDS concentration.
%
Moreover, in this range, the surfactant aggregate diffusion coefficient and the peptide diffusion coefficient coincide. This is a self-consistency check that gives confidence in the linear two species model and the results obtained. 

\begin{figure}[htp]
\centering
\includegraphics[scale=0.40]{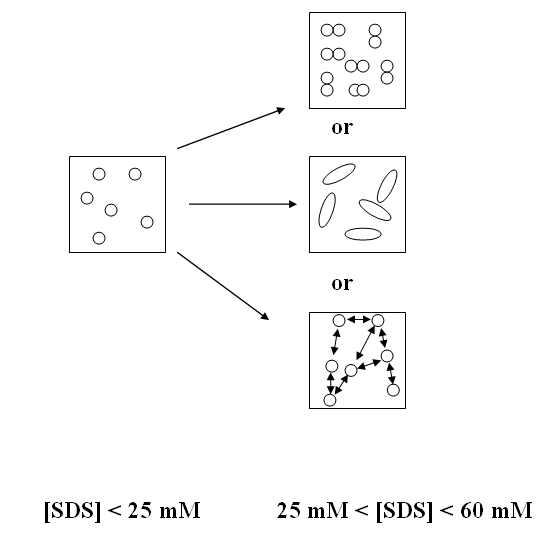}
\caption{A schematic diagram showing each peptide-surfactant complex as a single isolated complex (left, isolated circles) at low SDS concentrations. Results at intermediate SDS concentrations are consistent with either two peptides in each complex schematically represented by two circles (top right), highly anisotropic complexes (right middle), or long-range hydrodynamic interactions represented by arrows between complexes (right bottom).}
\label{AggregatesFormation}
\end{figure}

At intermediate SDS concentrations, the apparent hydrodynamic size increases from 5.5 nm at 25 mM to 10 nm at 60 mM (figure~\ref{Gad2SDSApparentRadiusGyration.eps}). This increase in the apparent hydrodynamic size might either reflect a true increase in aggregate size, or it might indicate the existence of hydrodynamic interactions between complexes. Given that the ratio of SDS to GAD-2 molecules in a complex is not changing by much, i.e. $r \approx R$ (figure~\ref{SDSAggregationGad2.eps}), an increase in the mean aggregate size might arise from an increase in the average number of peptides in one complex from 1 (at 25 mM) to approximately 2 (at 60 mM).
A third possibility is that such an increase in hydrodynamic radius arises from a change in shape (for example from spherical to oblate or prolate)~\citep{Bloomfield2000}. However, in order to account for a factor two increase, one would need to have a rather spectacular shape change with a formation of very anisotropic complexes with an approximately 20:1 axial ratio. These three possibilities - an increase in number of peptides in a complex, long-range interactions between complexes, or a dramatic change in complex shape - are depicted in figure~\ref{AggregatesFormation}. As noted
by~\citep{Zhou2008,Zhou2009}, a deeper understanding of role of electrostatic and hydrodynamic interactions is needed in the study of macromolecular crowding, and this needs to be studied further.

There is a distinct difference in the behavior of large peptides ($\mathrm{M_w^{peptide}} > \mathrm{M_w^{surfactant}}$) and small dipeptides ($\mathrm{M_w^{peptide}} \approx \mathrm{M_w^{surfactant}}$). The small dipeptides (Ala-Gly and Tyr-Leu) hardly affect the SDS diffusion coefficient (figure~\ref{SDSDiffusionForDifferentPeptides}). This indicates that the dipeptides behave just as the surfactant does: i.e. rapidly exchanging between aggregate and free state. For large peptides such as GAD-2, on the other hand, rapid exchange between free and aggregate state is practically impossible. This is because the surfactant molecules form micellar-like aggregates along the peptide chain, consistent with a bead-on-a-chain picture \citep{s5,PaolaRoscigno2003} for large-molecule aggregates. We therefore expect the approach outlined in this work to be valid generally for large hydrophobic peptides.

In conclusion, some recommendations are suggested in order to study peptides in membrane-mimic environments. All our results consistently show that measurements should be made in the regime where 
a two-species model is clearly valid, with the size of both free monomer and aggregate being independent of the surfactant concentration: this concentration is about 60 mM for pure SDS solutions. 
For peptide-SDS solutions, the true hydrodynamic size of the peptide-SDS complex is not necessarily constant
even at intermediate concentrations less than 60 mM, and the concentration dependence of the hydrodynamic radius can still not be ignored. The only unambiguous concentration-independent statements can be made at low concentrations: in this system, this is below [SDS]= 25 mM.

\section{Acknowledgments}
All the authors acknowledge financial support from the National Science and Engineering Research Council of Canada (NSERC). We also acknowledge useful suggestions from Carl Michal (University of British Coloumbia) and Ivan Saika-Voivod (Memorial University of Newfoundland).

\begin{thebibliography}{54}
\providecommand{\natexlab}[1]{#1}
\providecommand{\url}[1]{{#1}}
\providecommand{\urlprefix}{URL }
\expandafter\ifx\csname urlstyle\endcsname\relax
  \providecommand{\doi}[1]{DOI~\discretionary{}{}{}#1}\else
  \providecommand{\doi}{DOI~\discretionary{}{}{}\begingroup
  \urlstyle{rm}\Url}\fi
\providecommand{\eprint}[2][]{\url{#2}}

\bibitem[{Altieri et~al(1995)Altieri, Hinton, and Byrd}]{Byrd1995}
Altieri AS, Hinton DP, Byrd RA (1995) Association of biomolecular systems via
  pulsed field gradient {NMR} self-diffusion measurements. J Am Chem Soc
  117:7566--7561

\bibitem[{Andersson et~al(2004)Andersson, Almqvist, Hagn, and
  Maler}]{Andersson2004}
Andersson A, Almqvist J, Hagn F, Maler L (2004) Diffusion and dynamics of
  penetratin in different membrane mimicking media. Biochim Biophys Acta
  61:18--25

\bibitem[{Ando and Skolnick(2010)}]{TadashiAndo2010}
Ando T, Skolnick J (2010) Crowding and hydrodynamic interactions likely
  dominate in vivo macromolecular motion. {PNAS} 107:18,457--18,462

\bibitem[{Barhoum and Yethiraj(2010)}]{SulimanPolymer}
Barhoum S, Yethiraj A (2010) An {NMR} study of macromolecular aggregation in a
  model polymer-surfactant solution. J Chem Phys 132:1--9

\bibitem[{Batchelor(1976)}]{Batchelor1976}
Batchelor GK (1976) Brownian diffusion of particles with hydrodynamic
  interaction. J  Fluid Mech 74:1--29

\bibitem[{Begotka et~al(2006)Begotka, Hunsader, Oparaeche, Vincent, and
  Morris}]{Begotka2006}
Begotka BA, Hunsader JL, Oparaeche C, Vincent JK, Morris KF (2006) A pulsed
  field gradient {NMR} diffusion investigation of enkephalin peptide-sodium
  dodecyl sulfate micelle association. Magn Reson Chem 44:586--593

\bibitem[{Berr and Jones(1988)}]{Jones1988}
Berr SS, Jones RRM (1988) Effect of added sodium and lithium chlorides on
  intermicellar interactions and micellar size of aqueous dodecyl sulfate
  aggregates as determined by small-angle neutron scattering. Langmuir
  6:1247--1251

\bibitem[{Binks et~al(1989)Binks, Chatenay, Nicot, Urbach, and Waks}]{Waks1989}
Binks BP, Chatenay D, Nicot C, Urbach W, Waks M (1989) Structural parameters of
  the myelin transmembrane proteolipid in reverse micelles. Biophys J
  55:949--955

\bibitem[{Bloomfield(2000)}]{Bloomfield2000}
Bloomfield VA (2000) Survey of Biomolecular Hydrodynamics. In: Separations and
  Hydrodynamics.(Todd M. Schuster, editor). On-Line Biophysics Textbook,
  Biophysics society, www.biophysics.org

\bibitem[{Browne et~al(2011)Browne, Feng, Booth, and Rise}]{Rise2011}
Browne MJ, Feng CY, Booth V, Rise ML (2011) Characterization and expression
  studies of gaduscidin-1 and gaduscidin-2; paralogous antimicrobial
  peptide-like transcripts from atlantic cod (gadus morhua). Dev Comp Immunol
  35:399--408

\bibitem[{Buchko et~al(1998)Buchko, Rozek, Hoyt, Cushley, and
  Kennedy}]{Buchk1998}
Buchko GW, Rozek A, Hoyt DW, Cushley RJ, Kennedy MA (1998) The use of sodium
  dodecyl sulfate to model the apolipoprotein environment. evidence for peptide
  {SDS} complexes using pulsed-field-gradient {NMR} spectroscopy. Biochim
  Biophys Acta 1392:101--108

\bibitem[{Chari et~al(2004)Chari, Kowalczyk, and Lal}]{s5}
Chari K, Kowalczyk J, Lal J (2004) Conformation of poly(ethylene oxide) in
  polymer-surfactant aggregates. J Phys Chem B 108:2857--2861

\bibitem[{Chatterjee et~al({2004})Chatterjee, Majumder, and
  Mukhopadhyay}]{ChiradipChatterjee2004}
Chatterjee C, Majumder B, Mukhopadhyay C ({2004}) Pulsed-field gradient and
  saturation transfer difference {NMR} study of enkephalins in the ganglioside
  {GM1} micelle. J Phys Chem B {108}:{7430--7436}

\bibitem[{Chen et~al(1995)Chen, Wu, and Johnson}]{AidiChen1995}
Chen A, Wu D, Johnson CSJ (1995) Determination of the binding isotherm and size
  of the bovine serum albumin-sodium dodecyl sulfate complex by
  diffusion-ordered {2D} {NMR}. J Phys Chem 99:828--834

\bibitem[{Chinchar et~al(2004)Chinchar, Bryan, Silphadaung, Noga, Wade, and
  Rollins-Smith}]{Chinchar2004}
Chinchar V, Bryan L, Silphadaung U, Noga E, Wade D, Rollins-Smith L (2004)
  Inactivation of viruses infecting ectothermic animals by amphibian and
  piscine antimicrobial peptides. J Virol 323:268--275

\bibitem[{Cozzolino et~al(2008)Cozzolino, Sanna, and
  Valentini}]{SaraCozzolino2008}
Cozzolino S, Sanna MG, Valentini M (2008) Probing interactions by means of
  pulsed field gradient nuclear magnetic resonance spectroscopy. Magn Reson
  Chem 46:S16�S23

\bibitem[{Deaton et~al(2001)Deaton, Feyen, Nkulabi, and Morris}]{Deaton2001}
Deaton KR, Feyen EA, Nkulabi HJ, Morris KF (2001) Pulsed-field gradient {NMR}
  study of sodium dodecyl sulfate micelle-peptide association. Magn Reson Chem
  39:276--282

\bibitem[{Epand and Vogel(1999)}]{Epand1999}
Epand RM, Vogel HJ (1999) Diversity of antimicrobial peptides and their
  mechanisms of action. Biochim Biophys Acta 1462:11--28

\bibitem[{Fernandes et~al(2010)Fernandes, Ruangsri, and Kiron}]{Fernandes2010}
Fernandes JMO, Ruangsri J, Kiron V (2010) Atlantic cod piscidin and its
  diversification through positive selection. Public Library of Science 5:1--7

\bibitem[{Gao and Wong(1998)}]{Gao1998}
Gao X, Wong TC (1998) Studies of the binding and structure of
  adrenocorticotropin peptides in membrane mimics by {NMR} spectroscopy and
  pulsed-field gradient diffusion. Biophys J 75:1871--1888

\bibitem[{Gimel and Brown(1996)}]{Brown1996}
Gimel JC, Brown W (1996) A light scattering investigation of the sodium dodecyl
  sulfate-lysozyme system. J Chem Phys 104:8112--8117

\bibitem[{Hinton and Johnson(1994)}]{Johnson1994}
Hinton DP, Johnson CSJ (1994) Simultaneous measurement of vesicle diffusion
  coefficients and trapping efficiencies by means of diffusion ordered {2D NMR}
  spectroscopy. Chem Phys Lipids 69:175--178

\bibitem[{Hoskin and Ramamoorthy(2008)}]{Hoskin2008}
Hoskin DW, Ramamoorthy A (2008) Studies on anticancer activities of
  antimicrobial peptides. Biochim Biophys Acta 1778:357--375

\bibitem[{Iyota and Krastev(2009)}]{Krastev2009}
Iyota H, Krastev R (2009) Miscibility of sodium chloride and sodium dodecyl
  sulfate in the adsorbed film and aggregate. Colloid Polym Sci 287:425--433

\bibitem[{Jones et~al(1997)Jones, Wilkins, Smith, and Dobson}]{Dobson1997}
Jones JA, Wilkins DK, Smith LJ, Dobson CM (1997) Characterisation of protein
  unfolding by {NMR} diffusion measurements. J Biomol NMR 10:199--203

\bibitem[{Jones(2002)}]{jones}
Jones RAL (2002) Soft Condensed Matter, 1st edn. Oxford University Press Inc,
  New York

\bibitem[{Morein et~al(1996)Morein, Trouard, Hauksson, Rilfors, Arvidson, and
  Lindblom}]{SvenMorein1996}
Morein S, Trouard TP, Hauksson JB, Rilfors U, Arvidson G, Lindblom G (1996)
  Two-dimensional $^{1}${H}-{NMR} of transmembrane peptides from escherichia
  coli phosphatidylglycerophosphate synthase in micelles. Eur J Biochem
  241:489--497

\bibitem[{Morns and Johnson(1993)}]{Johnson1993}
Morns KF, Johnson CSJ (1993) Resolution of discrete and continuous molecular
  size distributions by means of diffusion-ordered {2D NMR} spectroscopy. J Am
  Chem Soc 115:4291--4299

\bibitem[{Morris and Johnson(1992)}]{Johnson1992}
Morris KF, Johnson CSJ (1992) Diffusion-ordered two-dimensional nuclear
  magnetic resonance spectroscopy. J Am Chem Soc 114:3139--3141

\bibitem[{Morris et~al(2005)Morris, Froberg, Becker, Almeida, Tarus, and
  Larive}]{Larive2005}
Morris KF, Froberg AL, Becker BA, Almeida VK, Tarus J, Larive CK (2005) Using
  {NMR} to develop insights into electrokinetic chromatography. Anal Chem
  77:254 A--263 A

\bibitem[{Nicolas(2009)}]{Nicolas2009}
Nicolas P (2009) Multifunctional host defense peptides: intracellular-targeting
  antimicrobial peptides. Federation of European Biochemical Societies
  276:6483--6496

\bibitem[{Orfi et~al(1998)Orfi, Lin, and Larive}]{Orfi1998}
Orfi L, Lin M, Larive CK (1998) Measurement of {SDS} micelle-peptide
  association using $^{1}${H} {NMR} chemical shift analysis and pulsed-field
  gradient nmr spectroscopy. J Anal Chem + 70:1339--1345

\bibitem[{Price(1997)}]{s1}
Price WS (1997) Pulsed-field gradient nuclear magnetic resonance as a tool for
  studying translational diffusion: Part {I}. basic theory. Concept Magnetic
  Res 9:299--336

\bibitem[{Qureshi and Goto(2012)}]{Goto2012}
Qureshi T, Goto NK (2012) Contemporary methods in structure determination of
  membrane proteins by solution {NMR}. Top Curr Chem 326:123--185

\bibitem[{Rege et~al(2007)Rege, Patel, Megeed, and Yarmush}]{Rege2007}
Rege K, Patel SJ, Megeed Z, Yarmush ML (2007) Amphipathic peptide-based fusion
  peptides and immunoconjugates for the targeted ablation of prostate cancer
  cells. Cancer Research Journal 67:6368--2375

\bibitem[{Romani et~al(2010)Romani, Marquezina, and Ito}]{AnaPaulaRomani2010}
Romani AP, Marquezina CA, Ito AS (2010) Fluorescence spectroscopy of small
  peptides interacting with microheterogeneous micelles. Int J Pharm
  383:154--156

\bibitem[{Roscigno et~al(2003)Roscigno, Asaro, Pellizer, Ortona, and
  Paduano}]{PaolaRoscigno2003}
Roscigno P, Asaro F, Pellizer G, Ortona O, Paduano L (2003) Complex formation
  between poly(vinylpyrrolidone) and sodium decyl sulfate studied through
  {NMR}. J Am Chem Soc 19:9639--9644

\bibitem[{Ruangsri et~al(2012)Ruangsri, Salger, Caipang, Kiron, and
  Fernandes}]{Ruangsri2012}
Ruangsri J, Salger SA, Caipang CM, Kiron V, Fernandes JM (2012) Differential
  expression and biological activity of two piscidin paralogues and a novel
  splice variant in atlantic cod ((g)adus morhua l.). Fish Shellfish Immun
  32:396--406

\bibitem[{Sammalkorpi et~al(2009)Sammalkorpi, Karttunen, and
  Haataja}]{Haataja2009}
Sammalkorpi M, Karttunen M, Haataja M (2009) Ionic surfactant aggregates in
  saline solutions: Sodium dodecyl sulfate ({SDS}) in the presence of excess
  sodium chloride ({NaCl}) or calcium chloride ($\mathrm{CaCl_2}$). J Phys Chem
  B 113:5863--5870

\bibitem[{Sanders and S\"{o}nnichsen(2006)}]{Charles2006}
Sanders CR, S\"{o}nnichsen F (2006) Solution {NMR} of membrane proteins:
  practice and challenges. Magn Reson Chem 44:s24--s40

\bibitem[{Sarker et~al(2011)Sarker, Rose, McDonald, Morrow, and
  Booth}]{Booth2011}
Sarker M, Rose J, McDonald M, Morrow MR, Booth V (2011) Modifications to
  surfactant protein b structure and lipid interactions under respiratory
  distress conditions: Consequences of tryptophan oxidation. J Biomol NMR
  50:25--36

\bibitem[{Schreiber et~al(2009)Schreiber, Haran, and Zhou}]{Zhou2009}
Schreiber G, Haran G, Zhou HX (2009) Fundamental aspects of protein-protein
  association kinetics. Chem Rev 109:839--860

\bibitem[{Soderman and Stilbs(1994)}]{Soderman1994}
Soderman O, Stilbs P (1994) {NMR} studies of complex surfactant systems. Prog
  Nucl Mag Res Sp 26:445--482

\bibitem[{Stilbs(1982)}]{Stilbs1982}
Stilbs P (1982) Fourier transform {NMR} pulsed-gradient spin-echo ({FT-PFGSE})
  self diffusion measurements of solubilization equilibria in sds solutions. J
  Colloid Interface Sci 87:385--394

\bibitem[{Stilbs(1983)}]{Stilbs1983}
Stilbs P (1983) A comparative study of micellar solubilization for combinations
  of surfactants and solubilizates using the fourier transform pulsed-gradient
  spin-echo {NMR} multicomponent self-diffusion technique. J Colloid Interface
  Sci 94:463--469

\bibitem[{Tulumello and Deber(2009)}]{Tulumello2009}
Tulumello DV, Deber CM (2009) {SDS} micelles as a membrane-mimetic environment
  for transmembrane segments. J Biochem 48:12,096--12,103

\bibitem[{Wang(1999)}]{Guangshun2006}
Wang G (1999) Structural biology of antimicrobial peptides by {NMR}
  spectroscopy. Curr Org Chem 10:569--581

\bibitem[{Wang(2008)}]{Guangshun2008}
Wang G (2008) {NMR} studies of a model antimicrobial peptide in the micelles of
  {SDS}, dodecylphosphocholine, or dioctanoylphosphatidylglycerol. The Open
  Magnetic Resonance Journal 1:9--15

\bibitem[{Whitehead et~al(2001)Whitehead, Jones, and
  Hicks}]{TracyWhitehead2001}
Whitehead TL, Jones LM, Hicks RP (2001) Effects of the incorporation of {CHAPS}
  into {SDS} micelles on neuropeptide-micelle binding: Separation of the role
  of electrostatic interactions from hydrophobic interactions. Biopolymers
  58:593--605

\bibitem[{Whitehead et~al(2004)Whitehead, Jones, and
  Hicks}]{TracyWhitehead2004}
Whitehead TL, Jones LM, Hicks RP (2004) {PFG-NMR} investigations of the binding
  of cationic neuropeptides to anionic and zwitterionic micelles. J Biomol
  Struct Dyn 21:567--576

\bibitem[{Wu et~al(1994)Wu, Chen, and Johnson}]{DWu1994}
Wu D, Chen A, Johnson CS (1994) An improved diffusion ordered-spectroscopy
  experiment incorporating bipolar-gradient pulses. J Magn Reson Ser A
  115:260--264

\bibitem[{Yu et~al(2006)Yu, Tan, Ho, Ding, and Wohland}]{LanlanYu2006}
Yu L, Tan M, Ho B, Ding JL, Wohland T (2006) Determination of critical micelle
  concentrations and aggregation numbers by fluorescence correlation
  spectroscopy: Aggregation of a lipopolysaccharide. Anal Chim Acta
  556:216--225

\bibitem[{Zasloff(2002)}]{Zasllof2002}
Zasloff M (2002) Antimicrobial peptides of multicellular organisms. Nature
  415:389--395

\bibitem[{Zhou et~al(2008)Zhou, Rivas, , and Minton}]{Zhou2008}
Zhou HX, Rivas G, , Minton AP (2008) Macromolecular crowding and confinement:
  biochemical, biophysical, and potential physiological consequences. Annu Rev
  Biophys 37:375--397

\end{thebibliography}
\end{document}